\documentclass[pra,twocolumn,amsmath,amssymb,floatfix,superscriptaddress]{revtex4}

\usepackage{amsmath}
\usepackage{graphicx}% Include figure files
\usepackage{epsfig}
\usepackage{bm}% bold math

\begin{document}

\title{Mapping of Coulomb gases and sine-Gordon models to statistics of random surfaces}
%New class of sine-Gordon models

\author{Adilet Imambekov}
\affiliation{Department of Physics, Harvard University, Cambridge,
Massachusetts 02138, USA} \affiliation{Department of Physics, Yale
University, New Haven, Connecticut 06520, USA}
\author{Vladimir Gritsev}
\affiliation{Department of Physics, Harvard University, Cambridge,
Massachusetts 02138, USA}
 \author{Eugene Demler}
\affiliation{Department of Physics, Harvard University, Cambridge,
Massachusetts 02138, USA}

\date{\today}
\begin{abstract}
We introduce a new class of sine-Gordon models, for which the
interaction term is present in a region different from the domain
over which the quadratic part is defined. We develop a {\it
nonperturbative} approach for calculating partition functions of
such models, which relies on mapping them to statistical
properties of random surfaces.  As a specific application of our
method, we consider the problem of calculating the amplitude of
interference fringes in experiments with two independent low
dimensional Bose gases. We calculate full distribution functions
of interference amplitude for one-dimensional and two-dimensional
gases with nonzero temperatures.
%We propose that distribution functions of interference fringe visibilities can be used for thermometry.
\end{abstract}

%%%%%%%%%%%%%%%%%%%%%%%%%%%%%%

\maketitle

\section{Introduction}

Sine-Gordon (SG) models and their Coulomb gas representations
appear as a low energy effective theory for many types of
physical systems. The so-called ``bulk'' SG model describes the
Berezinskii-Kosterlitz-Thouless (BKT) transition in
two-dimensional superfluids \cite{Berezinskii,KT,Tsvelik} and the
superfluid to insulator transition of Cooper pairs in a chain of
Josephson junctions \cite{junctions}. The so-called ``boundary''
SG model \cite{boundarySG} can be used to describe the
Chakravarty-Schmid transition in a single Josephson junction with
dissipation \cite{Chakravarty-Schmid} and a  quantum impurity
problem \cite{kanefisher} in one dimension (1D).
%and Kondo effect\cite{x}.
Powerful theoretical techniques have been developed for studying
such SG models including Bethe ansatz solutions \cite{SGBA},
renormalization group analysis (see e.g., Ref. \cite{Tsvelik}),
and functional renormalization group \cite{wegner}. Recent
theoretical work suggested that there is another class of SG
models which is important for several kinds of physical systems.
Such models can be described by the action
\begin{eqnarray}
S(g)=\pi K \int_\Omega \, (\vec \nabla  \phi )^2 dx\,d\tau  +
2\int_\omega g \, \cos{\left(2\pi\phi\right)} dx\,d\tau
\label{s_Gordon_finite},
\end{eqnarray}
where the interaction term $\cos{\left(2\pi\phi\right)}$ is
present in the spatial region $\omega$ (or space-time region for
quantum problems), which is only a part of the domain $\Omega$
over which the noninteracting part $(\vec \nabla \phi )^2$ is
defined. This should be contrasted to the bulk SG models, in
which the interaction term is present in the entire region
$\Omega$, and to the boundary SG models, in which the interaction
term is present on a line. We note that models with inhomogeneous
$g(x,\tau)$ can be considered using methods developed in the paper
as well, but for concreteness we will consider only constant $g.$
Model (\ref{s_Gordon_finite}) interpolates between the bulk and
boundary SG models, and we will refer to it as the interior
sine-Gordon (ISG) model.
%We point out that none of the standard
%theoretical approaches can be used to analyze ISG models.
%%%%%%%%%%%%%%%%%%%%%%%%%%%%%%%%%%%%%%%%%%%%%%%%%%%%%%%%%%%%%%%%%%
Here are a few examples of physical systems that can be described
by such models. The first example is the problem of ''interwire
coherence`` \cite{AdyStern} in which wires are brought together
over a finite length $l$ and separated on both ends. The
$\cos$-term describes the correlated umklapp electron scattering
in the two wires ($+2k_F$ scattering in one wire accompanied by
$-2k_F$ scattering in the other wire). The quantum space-time
action of this system has the form of Eq. (\ref{s_Gordon_finite})
with $\Omega = [-\infty,\infty]_x \times [0,\beta]_\tau$ and
$\omega = [-l/2,l/2]_x \times [0,\beta]_\tau$.
%$x \in [-\infty,\infty]$, $\tau \in [0,\beta]$ and
%$\omega$ $x \in [-l/2,l/2]$, $\tau \in [0,\beta]$.
Another example comes from a system of quantum particles in one
dimension (e.g., electrons or Cooper pairs in a wire, or
ultracold atoms in a weak optical trap) with a periodic potential
present in a finite region of the system. The $\cos$-term comes
from the umklapp scattering on the external potential and is
limited to the finite region in the interior of the system. The
third example is the problem of distribution functions of
interference fringe amplitudes (DFIFA) for a pair of independent
low-dimensional condensates \cite{Schumm,Hofferberth,Hadzibabic}.
Individual moments of the distribution function can be represented
as a microcanonical partition function of Coulomb gases
\cite{pnas,Gritsev,fcslong}, with positions of Coulomb charges
restricted to the part of the system from which interference
patterns have been extracted. The latter is typically smaller
than the total system size. For example, in the case of large 2D
condensates we have $\Omega = [-\infty,\infty]_x \times
[-\infty,\infty]_\tau$, and when the interference pattern is
extracted from the area $l \times l$, we have $\omega=
[-l/2,l/2]_x \times [-l/2,l/2]_\tau$. The relation between the
DFIFA and the partition function of the ISG model will be outlined
below and has also been discussed in Refs.
\cite{pnas,Gritsev,fcslong}.

In this paper we develop a nonperturbative approach to
calculating partition functions of a wide class of SG models and
Coulomb gases, which relies on the mapping of their partition
functions  to certain problems of statistics of random surfaces.
We point out that our method does not rely on the existence of
the exact solutions of SG models, but uses the structure of the
multi-point correlation functions in the absence of interactions.
We also note that a suitable extension of our method can be used
to compute correlation functions of SG models in equilibrium and
non-equilibrium situations. The particular strength of our
approach is that it can be applied to study ISG models, which
{\it cannot be analyzed by other theoretical methods}. As a
concrete application of our method we calculate DFIFA for both 1D
and 2D condensates. We point out that earlier theoretical work on
interference experiments focused on 1D systems with periodic
boundary conditions (PBC) \cite{Gritsev}. While these boundary
conditions are extremely artificial from the point of view of
realistic experiments, they allow one to relate the DFIFA to the
quantum impurity problem \cite{kanefisher} and use certain exact
results about the latter \cite{BLZ1-3}. Methods used in Ref.
\cite{Gritsev} can not be generalized either to the realistic
case of open boundary conditions (OBC) (i.e.,  interference
patterns extracted from the interior of a large system) or 2D
systems, but these cases can be analyzed using the method
discussed in this paper.  We emphasize, however, that the main
goal of this paper is to introduce an approach to the analysis of
SG models and the problem of DFIFA is just one example that
illustrates the power of the new method.

\section{Mapping}

The partition function corresponding to the action
(\ref{s_Gordon_finite}) is given by $Z(g)=\int {\cal D}\phi
e^{-S(g)}/\int {\cal D}\phi e^{-S(0)}$. By expanding $Z(g)$ in
powers of $g$ we arrive at the grand canonical partition function of
the Coulomb gas \cite{ChuiLee}
\begin{widetext}
\begin{equation}
Z(g)=\sum_{n=0}^{n=\infty}\frac{g^{2n}}{(n!)^2}Z_{2n} ,
\;\mbox{where}\; Z_{2n}=
\underset{\omega}{\int}...\underset{\omega}{\int} d^2 \vec
u_1...d^2 \vec v_n e^{\frac1K\left(\underset{i<j}{\sum}G(\vec
u_i,\vec u_j) + \underset{i<j}{\sum}G(\vec v_i ,\vec v_j)-
\underset{ij}{\sum}G(\vec u_i,\vec v_j) \right)}. \label{Z2n}
\end{equation}
\end{widetext} Here, $Z_{2n}$ is a microcanonical partition function
of a classical two-component neutral Coloumb gas of $2n$
particles, and $G(\vec x,\vec y)$ is an interaction potential,
which is proportional to Green's function of the Laplace operator
on $\Omega.$ The most familiar case is when $\Omega =
[-\infty,\infty]_x \times [-\infty,\infty]_\tau$ and $G(\vec
x,\vec y)=\ln{|\vec x-\vec y|}$.

To evaluate $Z(g)$ nonperturbatively in $g,$ we introduce an
auxiliary function $W(\alpha)$, which should be understood as a
certain distribution function, and is defined in such a way that
its $n\mbox{th}$ moment equals $Z_{2n}.$
\begin{eqnarray}
Z_{2n}=\int W(\alpha) \alpha^n d\alpha.\label{walpha}
\end{eqnarray}
One can use the Hankel transformation \cite{hankel} to compute
$Z(g)$ from $W(\alpha)$ as \begin{equation} Z(g)=\int_0^\infty
W(\alpha)I_0(2g\sqrt{\alpha}) d\alpha. \label{hankeleq}
\end{equation} This equation
equation can be verified using the Taylor expansion of the
modified Bessel function $I_0(x)$ and Eq. (\ref{walpha}).
Formulation of the auxiliary ''problem of moments`` allows one to
avoid calculating $Z(g)$ order by order, and can be viewed as a
tool to sum the perturbation series in Eq.~(\ref{Z2n}) to all
orders.

Function $G(\vec x,\vec y)$ is real and symmetric, so it can be diagonalized on
$\omega$ by solving the eigenvalue equations
\begin{eqnarray}
 \int_{\omega} G(\vec x,\vec y) \Psi_{f}(\vec y) d^2\vec y=G(f)\Psi_{f}(\vec x).  \label{inteqs}
\end{eqnarray}
Here $f$ is an integer index, which goes from $1$ to $\infty.$
$\Psi_{f}(\vec x)$ can be chosen to be real and normalized
according to $\int_{\omega}\Psi_{f}(\vec x)\Psi_{k}(\vec x)
d^2\vec x=\delta(f,k).$ Then, $G(\vec x,\vec y)$ is given by $
G(\vec x,\vec y)=\sum_{f=1}^{f=\infty}G(f) \Psi_{f}(\vec x)
\Psi_{f}(\vec y). $ Such decomposition  is similar to the
diagonalization of a symmetric matrix using its eigenvectors and
eigenvalues. We have
\begin{widetext}
\begin{eqnarray}
Z_{2n}= \int_{\omega}...\int_{\omega} d^2\vec u_1 ... d^2\vec v_n
e^{\sum_f\frac{
G(f)}{2K}\left[\left(\sum_{i=1}^{i=n}\Psi_{f}(\vec u_i)-\Psi_{f}(\vec v_i)\right)^2
-\sum_{i=1}^{i=n}\left(\Psi_{f}(\vec u_i)^2+\Psi_{f}(\vec v_i)^2
\right)\right]}=
\nonumber\\
\int_{\omega}...\int_{\omega} d^2\vec u_1 ...
d^2\vec v_n\prod_{f=1}^{f=\infty} \frac{ \int_{-\infty}^{\infty}d
t_{f}e^{-\frac{t_{f}^2}2} e^{\sum_{i}
t_f\sqrt{\frac{G(f)}{K}}\left(\Psi_{f}(\vec u_i)-\Psi_{f}(\vec v_i)
\right)-\frac{G(f)}{2K}\left(\Psi_{f}(\vec u_i)^2+\Psi_{f}(\vec v_i)^2
\right)}} {\sqrt{2\pi}}. \label{decoupling}
\end{eqnarray}
\end{widetext}
To go from the first to the second line in Eq. (\ref{decoupling})
we introduced the Hubbard-Stratonovich variables $t_f.$
Integration over $d^2\vec u_1 ... d^2\vec v_n,$ is now
straightforward since all $\vec u-$ and $\vec v-$ integrals are
identical.
\begin{eqnarray}
Z_{2n}=\left(\prod_{f=1}^{f=\infty}\frac{\int_{-\infty}^{\infty}
e^{-\frac{t_{f}^2}2}d
t_{f}}{\sqrt{2\pi}}\right)g(\{t_f\})^ng(\{-t_f\})^n, \label{10}
\end{eqnarray}
where
\begin{eqnarray}
g(\{t_f\})=\int_{\omega}d\vec x\; e^{\sum_f
t_f\sqrt{\frac{G(f)}{K}}\Psi_{f}(\vec x)-\frac{G(f)}{2K}\Psi_{f}(\vec x)^2}.\label{gt}
\end{eqnarray}
%If all eigenvalues $G(f)$ are negative, then
%\begin{eqnarray}
%g(-\{t_f\})= g(\{t_f\})^*, g(\{t_f\})g(\{-t_f\})=|g(\{t_f\})|^2.\nonumber
%\end{eqnarray}
From a comparison of Eqs. (\ref{walpha}) and (\ref{10}) we obtain
\begin{eqnarray}
W(\alpha)=\prod_{f=1}^{f=\infty}\frac{\int_{-\infty}^{\infty}
e^{-\frac{t_{f}^2}2}d
t_{f}}{\sqrt{2\pi}}\delta\left[\alpha-g(\{t_f\})g(\{-t_f\})\right].
\label{Walpha}
\end{eqnarray}
Equations (\ref{gt}) and (\ref{Walpha}) have a simple physical
interpretation. Consider $\Psi_{f}(\vec x)$ to be the eigenmodes
of the surface vibrations, $t_f$ the fluctuating mode amplitudes,
and $|G(f)|$ the noise power.  Infinite dimensional integral over
$\{t_f\}$ variables can be understood as an averaging over
fluctuations of the surface. For a particular realization of
noise variables $\{t_f\},$ a complex valued surface coordinate at
point $\vec x$ is given by $ h(\vec x;\{t_f\})= \sum_f t_f
\Psi_{f}(\vec x)\sqrt{G(f)/K}-G(f)\Psi_{f}(\vec x)^2/2K. $ For
each realization of a random surface $\{t_f\},$ $g(\{t_f\})$ is
obtained as an integral (\ref{gt}), which can also be written as
$g(\{t_f\})=\int_\omega d^2\vec x e^{h(\vec x;\{t_f\})}.$ Hence
Eq.~(\ref{Walpha}) can be interpreted as the mapping between the
partition function of the SG model (\ref{s_Gordon_finite}) and the
statistics of random surfaces subject to classical noise. This
mapping is the central result of this paper.

In general, $g(\{t_f\})$ is a complex number, thus in general,
$\alpha$ is defined on a complex plane. Simplifications occur: if
all eigenvalues $G(f)$ are negative, then $ g(\{-t_f\})=
g(\{t_f\})^*, g(\{t_f\})g(\{-t_f\})=|g(\{t_f\})|^2, $ and $\alpha$
is always real and positive.   Function $W(\alpha)$ can be
computed efficiently using Eq.~(\ref{Walpha}) and Monte Carlo
simulations. In the first step one solves integral
equations~(\ref{inteqs}) numerically to obtain eigenfunctions
$\Psi_{f}(\vec x)$ and eigenvectors $G(f).$ Then one samples
random numbers $\{t_f\}$ from the Gaussian ensemble, and plots the
histogram of the results for $g(\{t_f\})g(\{-t_f\}).$ Each point
on a histogram requires a computation of only two integrals, and
$W(\alpha)$ can be evaluated to arbitrary precision.

 Note, that this simple numerical evaluation of  $W(\alpha)$
allows one to extract $Z(g)$ for {\it all} values of $g$ using Eq.
(\ref{hankeleq}). While conventional large-scale Monte Carlo
simulations \cite{SGMC1,SGMC2} can be used to extract properties
of SG models, they require separate simulation for each value of
$g.$ Application of such methods to the calculation of DFIFA
described below would also require analytic continuation of
numerical results to imaginary values of $g$ \cite{Gritsev}, which
is numerically unstable. In addition, the mapping in
Eqs.~(\ref{gt}) and (\ref{Walpha}) does not only simplify
numerical simulations, but can also be used to obtain analytical
results (see below).

%\begin{figure}
%\includegraphics[scale=0.7,angle=0,bb=50 10 200 200]{setupnew.eps}
%%\psfig{file=setup.eps}
%\caption{\label{setup}
% Simplified
%setup of interference experiments with 1D Bose liquids (see e.g.
%Ref. \cite{Schumm,Hofferberth}). Two parallel condensates are
%extended in the $x$ direction. After atoms are released from the
%trap,  clouds are imaged by the laser beam propagating along the $z$
%axis. Meandering structure of the interference pattern arises from
%phase fluctuations along the condensates. The net interference
%amplitude $A$ is defined from the density integrated along the
%section of length $L.$  For the 2D setup is analogous, see e.g.
%Fig.~2 of Ref. \cite{pnas}.}
%\end{figure}

\section{Applications to interference of low-dimensional gases}

\begin{figure}
\psfig{file=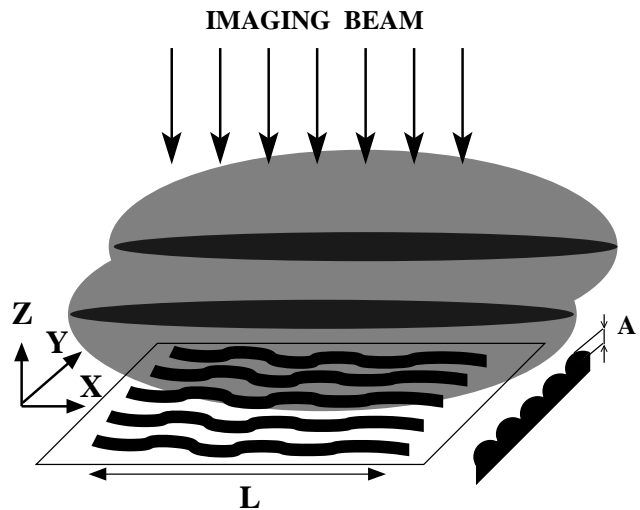} \caption{\label{setup}
 Simplified
setup of interference experiments with 1D Bose liquids (see, e.g.,
Refs. \cite{Schumm,Hofferberth}). Two parallel condensates are
extended in the $x$ direction. After atoms are released from the
trap,  clouds are imaged by the laser beam propagating along the
$z$ axis. Meandering structure of the interference pattern arises
from phase fluctuations along the condensates. The net
interference amplitude $A$ is defined from the density integrated
along the section of length $L.$  For the analogous 2D setup, see,
e.g., Fig.~2 of Ref. \cite{pnas}.}
\end{figure}

We now apply a general formalism developed to a particular problem
of the interference of low-dimensional Bose gases
\cite{Schumm,Hofferberth,Hadzibabic,pnas,Gritsev,fcslong}. Typical
experimental setup for interference of low-dimensional gases with
open boundary conditions (OBC) is shown in Fig. \ref{setup}.
%%, and corresponds to experimental
%%realization of Refs. \cite{Hadzibabic,Schumm,Hofferberth}.
Two parallel condensates are extended in the $x$ direction. After
atoms are released from the trap,  clouds expand predominantly in
the transverse direction. After sufficient time of flight clouds
overlap, and the laser beam propagating along the $z$ axis takes
an absorption image. Fluctuations of the relative phase result in
fluctuations of the minima positions for different $x.$ For each
$y,$ the image can be integrated along the $x$ direction to obtain
the integrated fringe amplitude $A.$ One experimental image can be
used to extract information for different values of $L.$ Many
images are still required to obtain distribution functions for
each $L.$

For two identical 1D clouds  higher moments of the fringe
amplitude $A$ can be written as \cite{pnas}
\begin{eqnarray}
\langle |A|^{2n}\rangle =A_0^{2n}  Z_{2n}, \; \mbox{where} \;
A_0=\sqrt{ C \rho^2 \xi_h^{1/K} L^{2-1/K}}.\label{A0}
\end{eqnarray}
Here $\rho$ is the density, $L$ is the imaging length, $\xi_h$ is
the healing length, and $C$ is a constant of the order of unity.
For OBC  $Z_{2n}$ is given \cite{pnas} by  Eq. (\ref{Z2n}) with
$\Omega=[-\infty,\infty]_x \times[0,\beta]_\tau$ (and periodic
boundary conditions in $\tau$) and $\omega=[0,1]_x$ (and fixed
$\tau$).  Equation (\ref{A0}) has been derived neglecting the shot
noise, which arises due to a finite number of particles in the
interfering clouds \cite{shotnoise,fcslong}. In what follows, we
will be interested in the distribution functions $W(\alpha)$ of a
positive variable $\alpha=|A|^{2}/A_0^2$ defined by Eq.
(\ref{walpha}), or of its normalized version $\tilde
\alpha=|A|^{2}/\langle A^2\rangle,$ defined by
$Z_{2n}/Z_2^n=\int_0^\infty \tilde W(\tilde \alpha)\tilde
\alpha^n d\tilde \alpha.$ For zero temperature $Z_{2n}$ depends
only on the Luttinger parameter $K,$ which describes
\cite{Haldane,Cazalilla} the long-distance behavior of boson
correlation functions, given by $\langle
a^{\dagger}(x)a(0)\rangle \sim \rho \left(\xi_h/x\right)^{1/2K}$.
For bosons $K$ ranges from $K=1$ (strong interactions) to
$K=\infty$ (weak interactions). For OBC and zero temperature
$G(\vec x,\vec y)$ equals
%\begin{widetext}
%\begin{eqnarray}\label{z2n}
% Z_{2n}=\int_{0}^1
%...\int_{0}^1 du_1 ...  dv_n
%\left|\frac{\underset{i<j}{\prod}|u_i -u_j| \underset{i<j}{\prod}|v_i
%-v_j|}{\underset{ij}{\prod} |u_i-v_j| } \right|^{\frac1K}
% =\int_0^1
%...\int_0^1 du_1 ... dv_n
%e^{\frac1K\left(\underset{i<j}{\sum}G(u_i,u_j) + \underset{i<j}{\sum}G(v_i
%,v_j)- \underset{ij}{\sum}G(u_i,v_j)  \right)}, \label{Z2n}
%\end{eqnarray}
%\end{widetext}
$G(\vec x,\vec y)=\log{|\vec x-\vec y|},$ while for PBC
considered in Ref.~\cite{Gritsev}, $G^{per}(\vec x,\vec
y)=\ln{\frac1\pi\sin{\pi|\vec x-\vec y|}}.$
 While $A_0$ depends on $L,$  for zero
temperature distribution $\tilde W(\tilde \alpha)$ does not depend
on $L,$ but depends only on $K.$ For nonzero temperature, $Z_{2n}$
depends on $K$ and the thermal length $\xi_T=\hbar v_s/(k_B T),$
where $v_s$ is the sound velocity: $ G(\vec x,\vec
y,\xi_T/L)=\ln{\left(\frac{\xi_T}{\pi L}\sinh{\frac{\pi|\vec
x-\vec y|L}{\xi_T}}\right)}. $

For 2D, one can use a similar approach to describe the contrast
distribution at finite temperature below the BKT transition. In
this case, correlation functions are given by $ \langle
a^{\dagger}(r)a(0)\rangle \sim \rho
\left(\xi_h/r\right)^{\eta(T)}, $ where $\eta(T)=m T/(2\pi\hbar^2
\rho_s(T))$ depends on the temperature and the superfluid density
$\rho_s(T).$ The BKT transition happens at the universal value
$\eta_c(T_c)=1/4.$ To keep a connection to the 1D case, we will
use $K=1/(2\eta(T)),$ and restrict our attention to $K>K_c=2.$
For 2D with the aspect ratio of the imaging area equal to unity,
$\vec u_i$ and $\vec v_i$ in Eq. (\ref{Z2n}) are defined on a
square $\omega=[0,1]_{x}\times[0,1]_{\tau}$ with $G(\vec x,\vec
y)=\ln{|\vec x-\vec y|}.$

\begin{figure}
\includegraphics[scale=0.8,angle=0,bb=50 10 200 150]{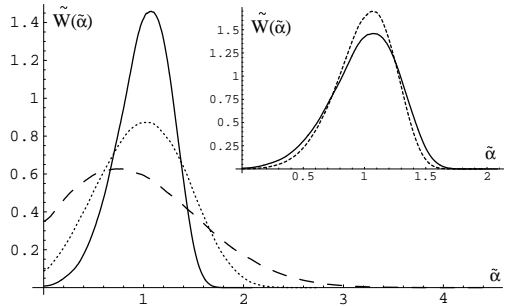}
\caption{\label{1dzeroT}Distribution functions of  the normalized
interference amplitude $\tilde W(\tilde \alpha)$ at $T=0$ for 1D
gases with open boundary conditions, shown for Luttinger parameters
$K=2\;\mbox{(dashed)}, K=3\;\mbox{(dotted)}, $ and $ K=5
\;\mbox{(solid)}.$ The inset shows a comparison between open (solid)
and periodic (dashed) boundary conditions for $K=5$.}
\end{figure}

In the limit $K\rightarrow \infty,$ one can expand the exponent of
Eq. (\ref{gt}) in the Taylor series. Then
$\alpha=g(\{t_f\})g(\{-t_f\})$ is linearly related to roughness,
or mean square fluctuation of the surface, as defined in
Ref.~\cite{1fnoise}. For PBC the noise has  a $1/f$ power
spectrum, since
$G^{per}(x,y)=\ln{\frac1\pi\sin{\pi|x-y|}}=-\ln{2\pi}-\sum_{f=1}^{f=\infty}(\cos{2\pi
f x}\cos{2\pi f y}+\sin{2\pi f x}\sin{2\pi f y})/f.$  This results
in the Gumbel statistics of the roughness \cite{Gritsev,1fnoise,
BertinClusel,fcslong}: $\tilde
W_{G}(\tilde\alpha)=K\;\exp(x-e^{x}),$ where
$x=K(\tilde{\alpha}-1)-\gamma$ and $\gamma=0.577$ is the Euler
constant. This provides the {\it analytical} proof of the
conjecture made in Ref. \cite{Gritsev}, that the distribution
function in this case is given by one of the extreme value
statistical distributions, the Gumbel function \cite{gumbelbook}.

 In what follows we perform
simulations of $W(\alpha)$ with up to $N=10^6-10^7$ realizations of
$\{t_f\}$ and smoothen the data. We use a finite value of $f_{max}$
and check for convergence with $f_{max},$ typically $\sim 30.$
$\langle \alpha\rangle $ is always kept within $1\%$ from its
expected value. For most of the presented results, all eigenvalues
$G(f)$ are negative, and Eq.  (\ref{Walpha}) can be directly
applied.
%since  $g(\{t_f\})g(\{-t_f\})=|g(\{t_f\})|^2$ is positive and real in these case.
Special care should be taken of the  1D case with nonzero
temperatures, since one of the eigenvalues  can be positive. This
situation can be handled by subtracting a sufficiently large
positive constant $C$ from $G(\vec x,\vec y,\xi_T/L),$ which
makes all eigenvalues negative. According to Eqs. (\ref{Z2n}) and
(\ref{walpha}), this leads to rescaling of $\alpha$ by a factor
$e^{-C/K},$ which can be easily taken into account.

\begin{figure}
\includegraphics[scale=0.8,angle=0,bb=50 10 200 150]{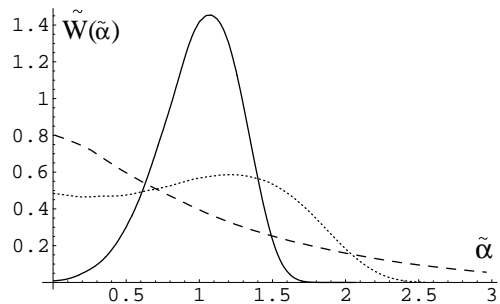}
\caption{\label{1dnztfig} Distribution functions of the normalized
interference amplitude $\tilde W(\tilde \alpha)$ for a 1D Bose gas
with open boundary conditions at nonzero temperature and $K=5.$
Different curves correspond to ratios $K \xi_T/L
=\infty\;\mbox{(solid)}, K \xi_T/L=1\;\mbox{(dotted)},$ and $K
\xi_T/L=0.25 \;\mbox{(dashed)}.$ $\xi_T$ is the thermal
correlation length, $K$ is the Luttinger parameter, and $L$ is
the imaging length.}
\end{figure}

\begin{figure}
\includegraphics[scale=0.8,angle=0,bb=50 10 200 150]{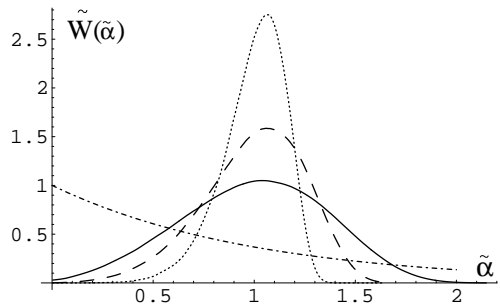}
\caption{\label{2dfig}Distribution functions of the normalized
interference amplitude $\tilde W(\tilde \alpha)$ for a
two-dimensional Bose gas with the aspect ratio of the imaging area
equal to unity and open boundary conditions. Temperature is below
the Berezinskii-Kosterlitz-Thouless (BKT) transition temperature.
Different curves correspond to $\eta(T)=\eta_c(T_c)=1/4$(the BKT
transition point, solid), $\eta(T)=1/6$ (dashed line), and
$\eta(T)=1/10$ (dotted line). Above the BKT transition temperature
the distribution function is Poissonian (dot-dashed line). }
\end{figure}

In Fig.~\ref{1dzeroT} we show distribution functions of the
normalized interference amplitude $\tilde W(\tilde \alpha)$ at
$T=0$ for 1D gases with OBC for various $K.$ The inset shows a
comparison between OBC and PBC for $K=5.$ In Fig. \ref{1dnztfig},
we show distribution functions of the normalized interference
amplitude for a 1D gas with OBC at nonzero temperature and $K=5.$
For $\xi_T K/L\ll1$ distribution is Poissonian \cite{pnas,
Gritsev, fcslong} and wide, while for $K\gg1$ and  $\xi_T
K/L\gg1$ it is very narrow. Evolution of the full distribution
function of the visibilities as $L$ is varied can be used to
precisely measure the thermal length $\xi_T,$  and to extract the
temperature.   As seen in Fig. \ref{1dnztfig}, at $T\neq 0$ the
distribution  function has characteristic features, i.e., it is
generally nonsymmetric and can have a {\it minimum}. These
features can be used to distinguish the intrinsic noise due to
fluctuations of the phase from technical noise. Finally, in
Fig.~\ref{2dfig} we show distribution functions of the normalized
interference amplitude for a 2D gas with an aspect ratio of
imaging area equal to unity and OBC below the BKT temperature.
Above the BKT temperature, distribution functions become
Poissonian for $L\gg \xi,$ where $\xi$ is the correlation length.
In 2D one cannot describe the crossover at $L\sim \xi$ similar to
1D, since the action which describes the fluctuations of the
phase is not quadratic in this region, and Eq. (\ref{Z2n}) does
not hold.

\section{Conclusions}

To summarize, we introduced a  class of sine-Gordon models, for
which an  interaction term is present in a spatial region
different from the domain over which the noninteracting part is
defined. We developed a general mapping of such sine-Gordon
models and related Coulomb gases to statistical properties of
random surfaces, which can be used to  calculate their partition
functions {\it nonperturbatively}.   As a specific application of
our approach, we considered interference experiments with two
independent low-dimensional Bose gases. We calculated full
distribution functions of the interference amplitude for 1D and
2D gases with open boundary conditions and nonzero temperatures.
Full distribution functions of interference fringe visibilities
can be used for thermometry.

We thank M.~Mueller and A.~Polkovnikov for useful discussions.
This work was partially supported by NSF Grant No. DMR-0705472,
MIT-Harvard CUA, and AFOSR. V.G. is also partially supported by
Swiss National Science Foundation, Grant No. PBFR2-110423.

\end{document}